\documentclass[a4paper,11pt]{article}
\pdfoutput=1 

\usepackage{jinstpub} 

\usepackage{lineno}

\title{Serial powering and signal integrity characterization for the TEPX detector for the Phase-2 CMS Inner Tracker}

\author[1]{A. C. Reimers\note{Speaker}}
\affiliation{Universit\"at Z\"urich, Switzerland}

\emailAdd{arne.reimers@cern.ch}

\abstract{The entire CMS silicon pixel detector will be replaced to operate at the High-Luminosity LHC. In this contribution, the new Tracker Endcap Pixel detector will be presented, focusing on a novel concept to provide both power and data connectivity to the modules through a disk-shaped PCB. As this part of the detector also features the longest serial powering chains in the CMS Phase-2 Inner Tracker, emphasis is put on serial powering results and the characterization of the signal integrity.}


\collaboration[c]{on behalf of the CMS Tracker Group}

\proceeding{Topical Workshop on Electronics for Particle Physics, TWEPP 2021\\20 - 24 September 2021}

\begin{document}
\maketitle
\flushbottom

\section{Introduction}
\label{sec:introduction}

The CERN LHC~\cite{Bruning:782076, Evans:2008zzb} is the world's most powerful and luminous particle collider to date.
After the end of the Run 3 data taking, the LHC will be upgraded to deliver instantaneous luminosities of up to $7.5\times 10^{34}\,\rm{cm}^{-2}\rm{s}^{-1}$, exceeding the initial design value by a factor of $7.5$~\cite{Aberle:2749422}.
The entire CMS detector will undergo substantial upgrades in order to fully exploit the physics potential of the collisions delivered while resisting the increased level of radiation.

\begin{figure}[hb]
\centering
\includegraphics[width=0.66\textwidth]{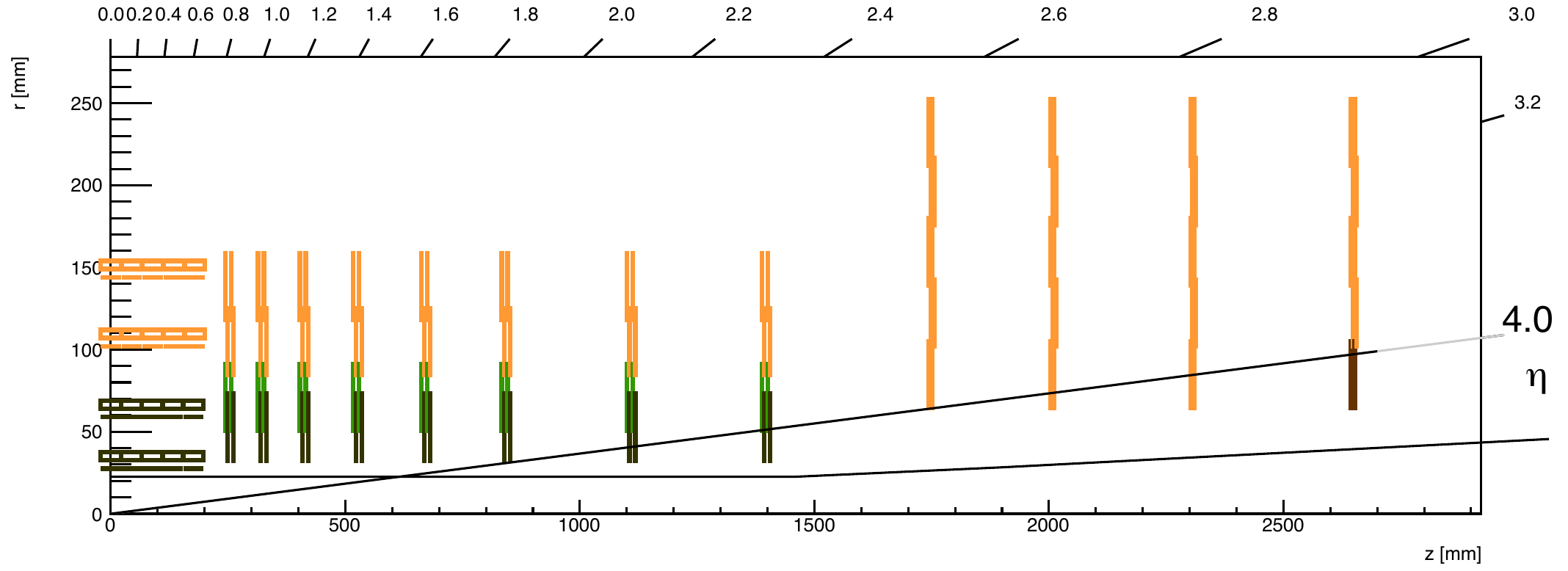} \hfill
\includegraphics[width=0.33\textwidth]{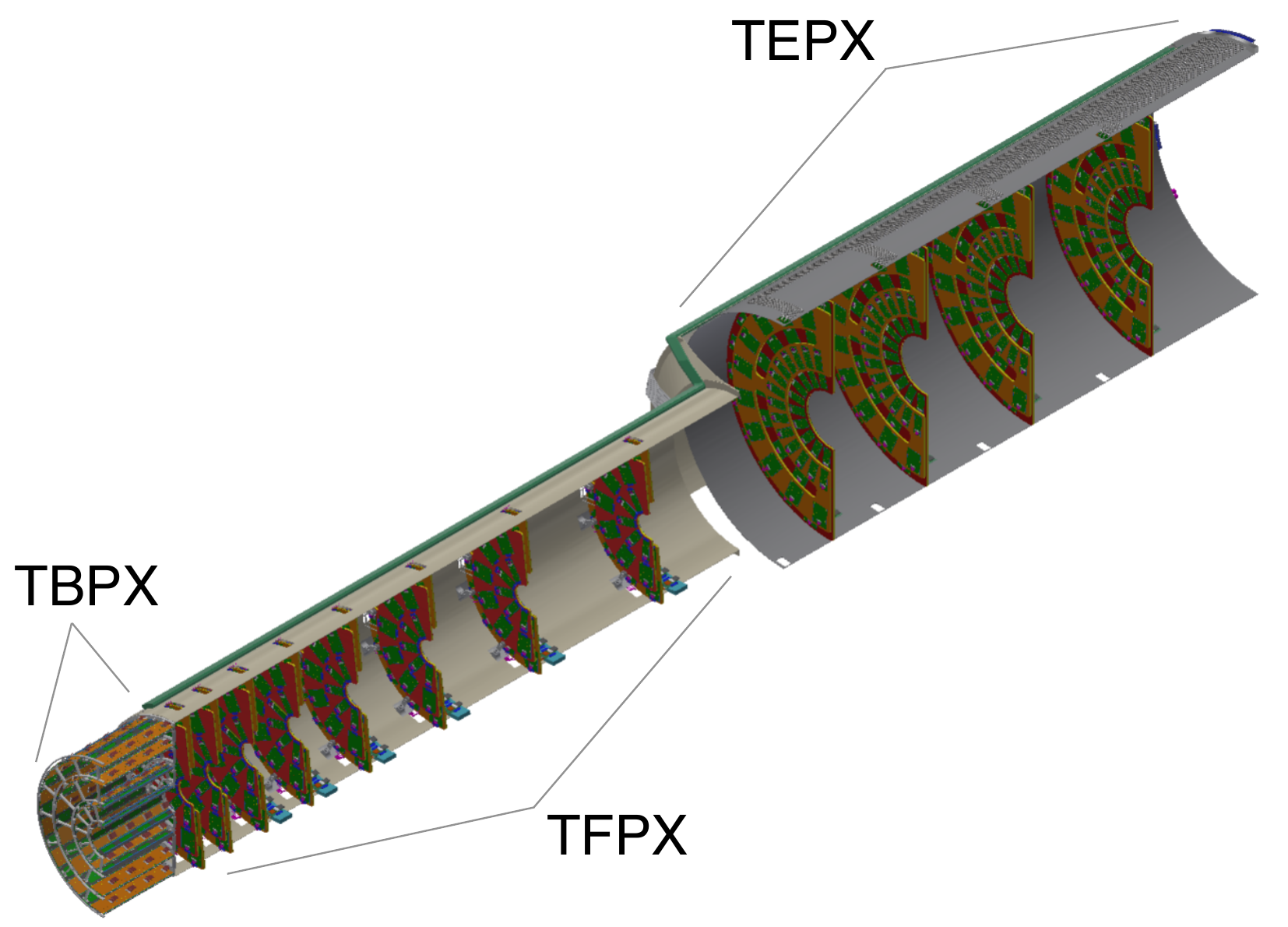}
\caption{Layout (left, updated from Ref.~\cite{CERN-LHCC-2017-009}) and schematic view~\cite{CERN-LHCC-2017-009} (right) of a quarter of the upgraded CMS Inner Tracker. The $z$-axis is aligned with the beam direction.}
\label{fig:ITSketches}
\end{figure}

The new pixel tracking system~\cite{CERN-LHCC-2017-009} will extend the coverage up to pseudorapidities of $|\eta| = 4.0$. The layout and a schematic view are shown in Figure~\ref{fig:ITSketches}.
Due to its proximity to the interaction point, enhanced radiation tolerance of the silicon sensor and readout chip are required.
A reduced material budget and a pixel size of $100\times 25$ or $50\times 50\,\mathrm{\mu}\rm{m}^2$ will be instrumental in ensuring stable tracking performance while maintaining a low occupancy despite the increased particle flux.
The new Inner Tracker is subdivided into the Tracker Barrel Pixel (TBPX), Tracker Forward Pixel (TFPX), and Tracker Endcap Pixel (TEPX) systems, which are visible in Figure~\ref{fig:ITSketches}.
The TEPX detector is presented in the following section, with particular emphasis on the characterization of the serial powering scheme and the electrical signal integrity of the TEPX readout chain.

\section{The Tracker Endcap Pixel Detector}
\label{sec:tepx}

\begin{figure}[t]
\centering
\includegraphics[width=0.685\textwidth]{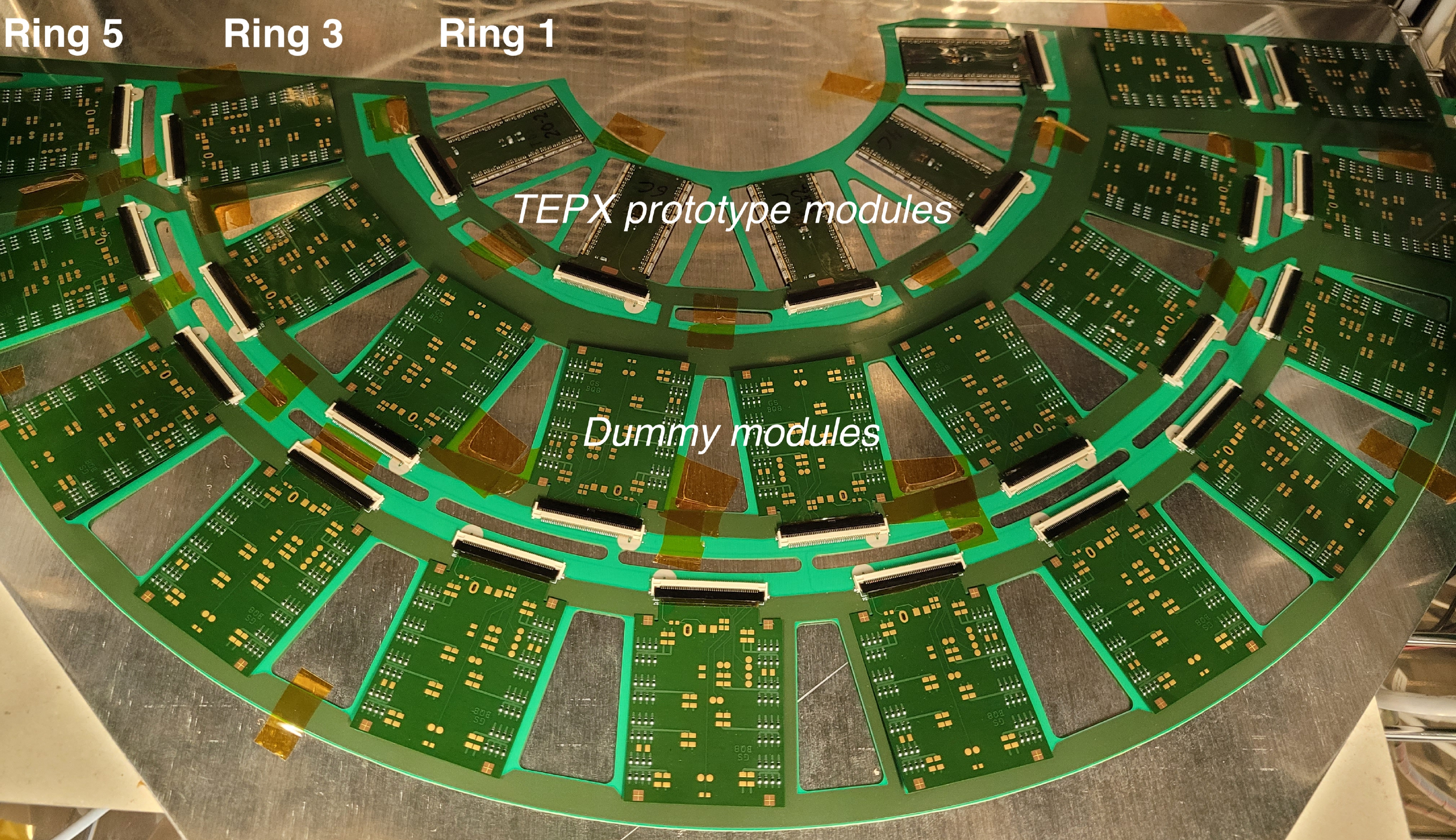} \hfill
\includegraphics[width=0.296\textwidth]{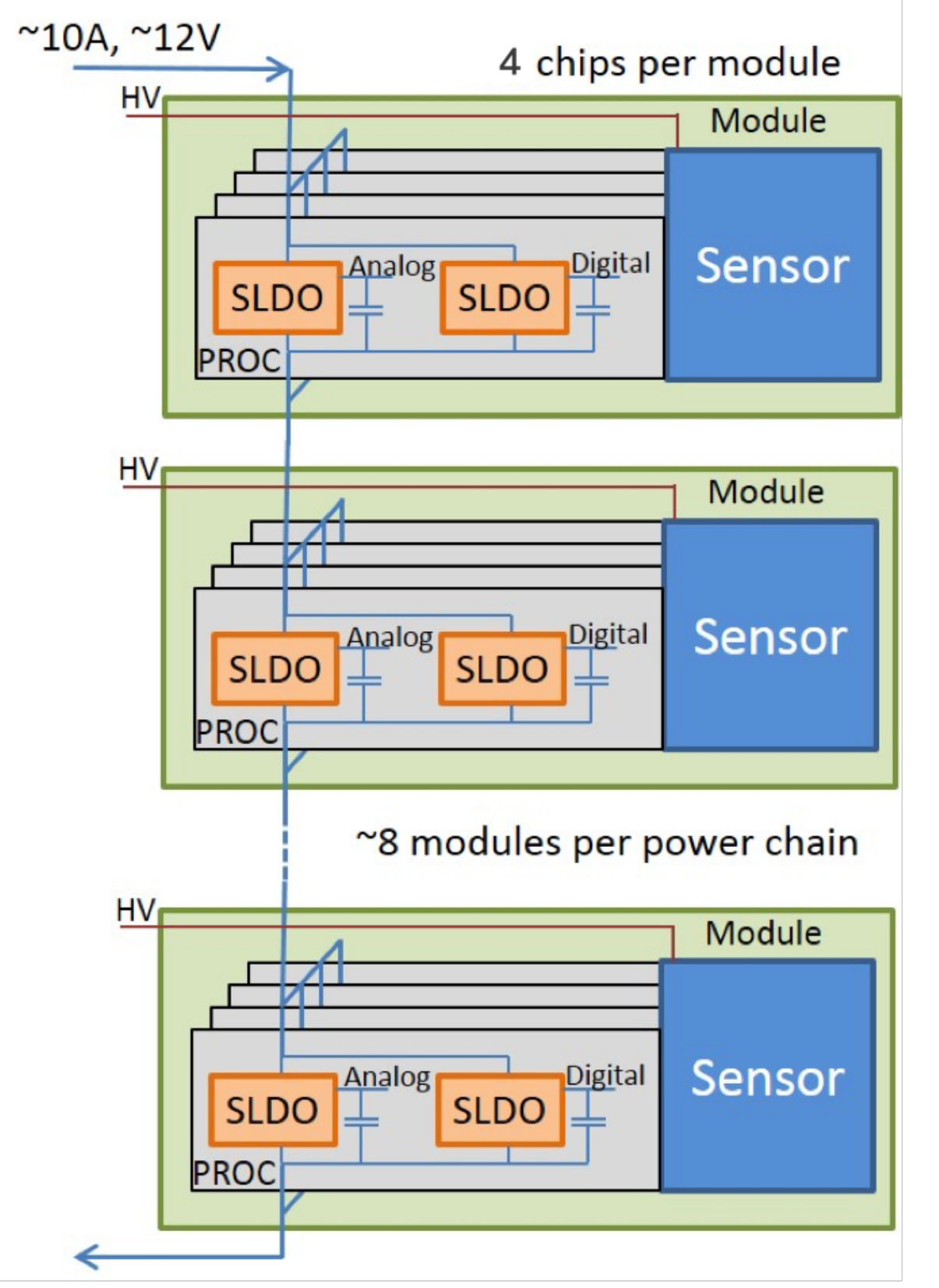}
\caption{Left: TEPX half-disk prototype equipped with TEPX prototype modules in ring 1 and dummy modules in rings 3 and 5. Right: Schematic of the serial powering chain used in the TEPX system~\cite{CERN-LHCC-2017-009}.}
\label{fig:DiskAndSP}
\end{figure}

The TEPX detector will consist of four double-disks in each of the CMS endcaps. In each disk, the TEPX modules will be arranged in five concentric rings, where the front face (towards the interaction point) of each disk will contain the odd-numbered rings and the back face will contain the even-numbered ones. Each disk is complemented by a second disk, thus forming a double-disk, in order to achieve hermetic coverage. Power and data connectivity will be provided to the modules through a five-layer Polyimide PCB of $400\,\mathrm{\mu}\rm{m}$ thickness with meshed input and return current layers. In Figure~\ref{fig:DiskAndSP} (left), the prototype of a TEPX half-disk with three rings is shown. The innermost ring 1 (R1) will hold five modules, the intermediate ring 3 nine, and the outermost ring~5 twelve TEPX modules. Modules in ring 1 are read out via three data lines per module because of the high expected data rate, while the modules in rings 3 and 5 are read out with a single line per module.

The TEPX modules will feature silicon sensors with a pixel size of $100\times 25$ or $50\times 50\,\mathrm{\mu}\rm{m}^2$, which are bump-bonded to four readout chips (ROCs or PROCs) per module. In the prototype modules studied here, the RD53A~\cite{Garcia-Sciveres:2287593} ROC is used. The ROCs are wire-bonded to an HDI with a data readout, bias voltage, and power connection, where the latter is regulated by two shunt-low-dropout (SLDO) regulators for each ROC, one for the analog and one for the digital supply line.
In the TEPX disk, all modules in a given ring are powered in series (serial powering, SP), while the current is shared between the four ROCs in a module. Conversely, the bias voltage (HV) is provided in parallel to all modules in the same ring. This powering scheme is illustrated in Figure~\ref{fig:DiskAndSP} (right).

The data lines of the TEPX disk prototype are connected to two mezzanine cards on an FC7~\cite{Pesaresi:2015qfa} readout board via flat flex and DisplayPort (DP) cables. Due to the half-disk prototype design, a maximum of three ROCs per module can be read from ring 1, one per data line. In the final TEPX detector, all four ROCs of each module will be read out. In standalone operation, a single TEPX prototype module is plugged into a flex-to-DP adapter, which is directly connected to the mezzanine card through a DP cable. All four ROCs of the module are read out simultaneously in this case.

\subsection{Serial Powering Characterization}
\label{subsec:sp}

In order to characterize the performance and reliablity of five TEPX modules powered in series in R1 of the TEPX half-disk prototype, the pixel thresholds are tuned and compared to results obtained in standalone operation of single modules. In both cases, the modules are operated with an input current of 6.8\,A at an ambient temperature of $-50^\circ\mathrm{C}$ inside a climatic chamber. The analogue and digital voltages of each ROC are tuned to 1.20 and 1.25\,V, respectively. Prototype modules with and without a silicon sensor are used, which are referred to as sensor modules and digital modules, respectively. A bias voltage of $-50\,\mathrm{V}$ is applied to deplete the sensor of the former.

In the procedure applied to tune the tresholds, initially unresponsive pixels of a given ROC are masked. After equalizing the channels of that ROC, noisy pixels are masked. The global threshold is adjusted to a target of 1500\,e before the channels are re-equalized. Finally, newly noisy pixels are masked and the efficiency as a function of the injected charge is measured for each pixel of a given ROC, from which the threshold and noise of each pixel is extracted.

\begin{figure}[t]
\centering
\includegraphics[width=0.39\textwidth,trim=0 0 0 50,clip]{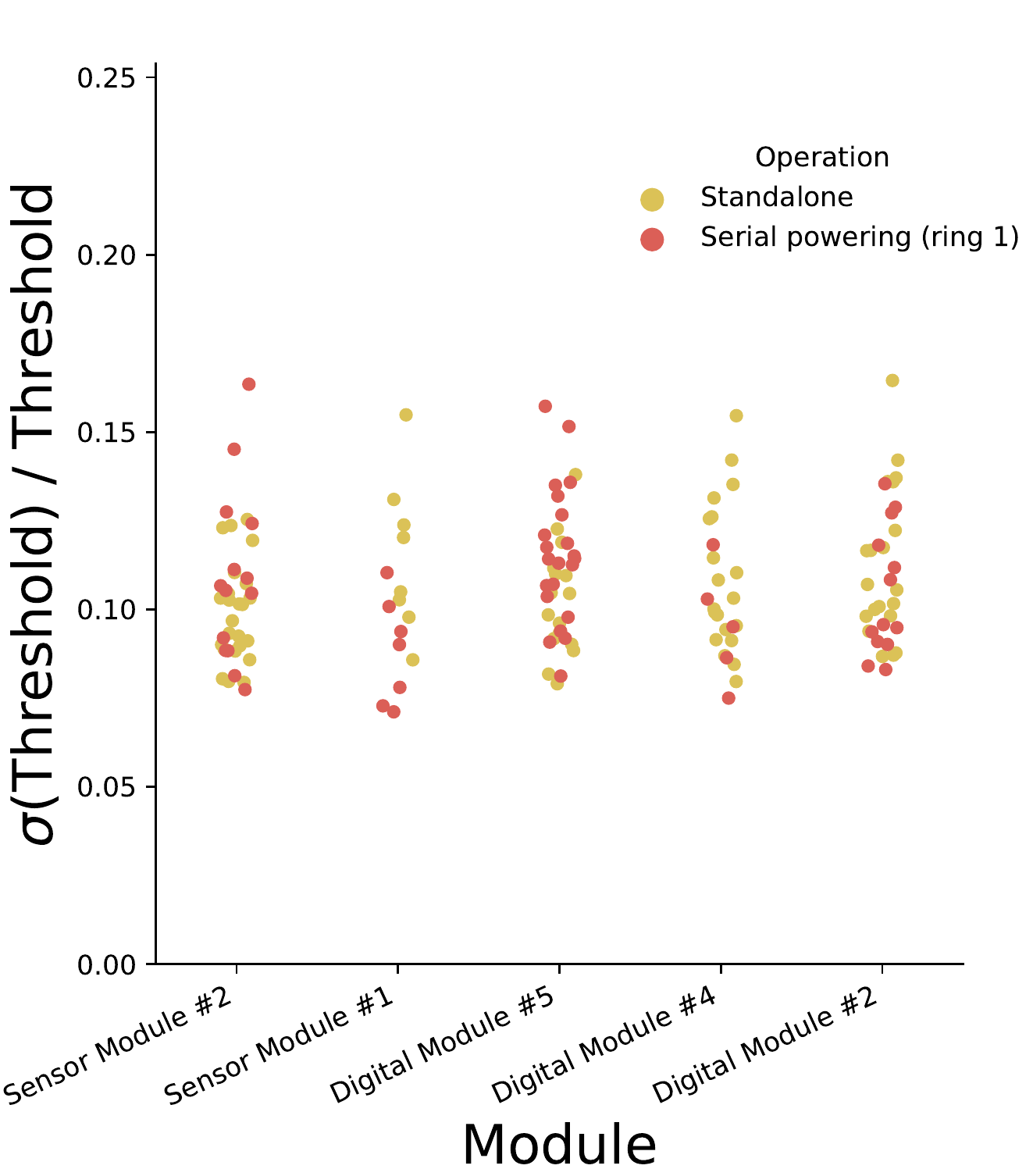} \hfill
\includegraphics[width=0.39\textwidth,trim=0 0 0 50,clip]{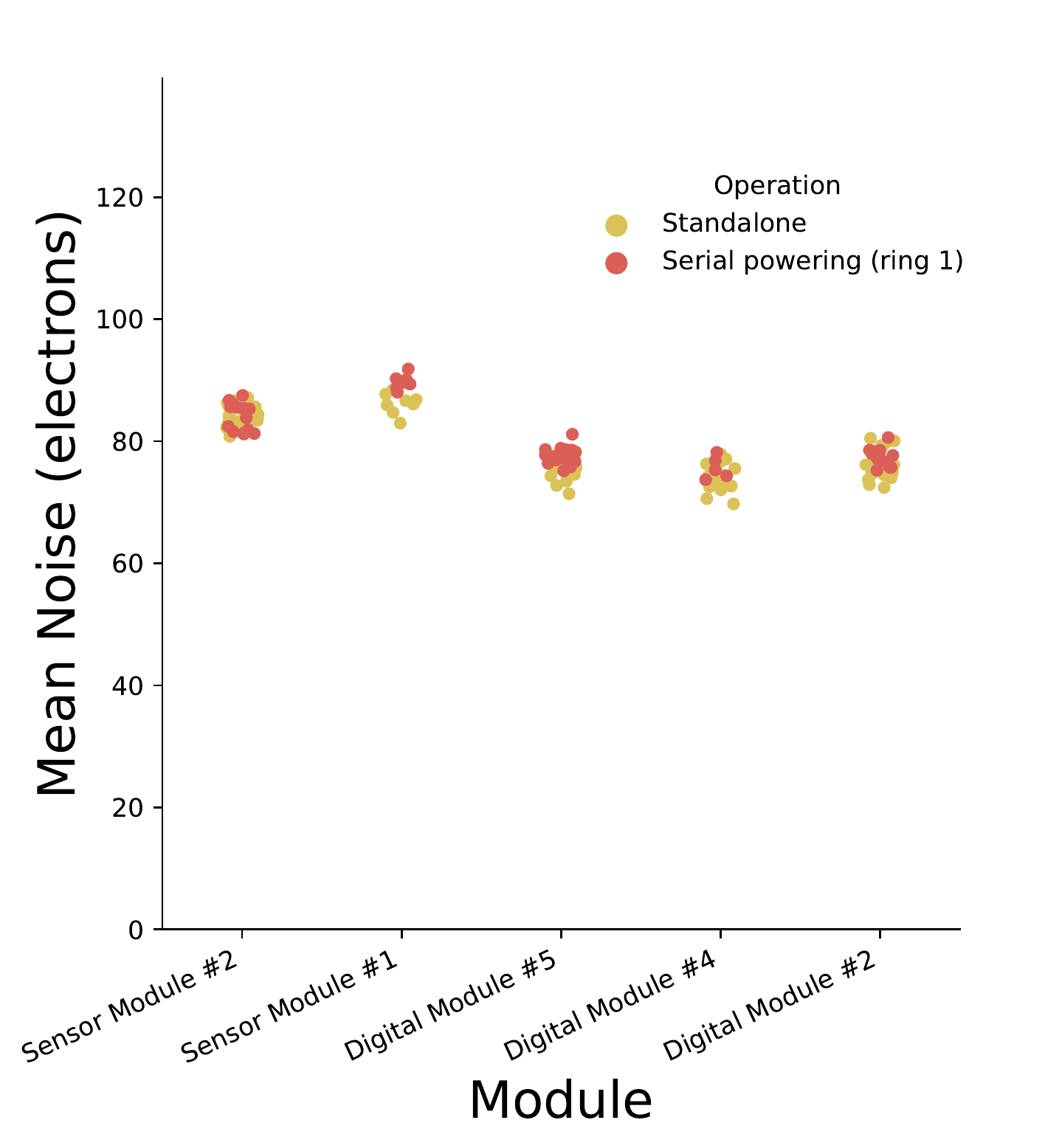}
\caption{Comparisons of the relative width of the threshold distribution (left) and the mean noise (right) between standalone (yellow) and serial powering operation (red) for different TEPX prototype modules ($x$-axis). Each point corresponds to a single measurement with a single ROC.}
\label{fig:ComparisonSABvsSP}
\end{figure}

In Figure~\ref{fig:ComparisonSABvsSP}, the relative width of the threshold distribution (left) and the mean noise (right) are compared for different modules and ROCs between standalone (yellow) and SP (red) operation. Both modes of operation yield very similar results with no systematic differences observed. It was verified that the number of masked noisy pixels as a function of the target threshold is consistent between SP and standalone operation and no additional noise is introduced by the SP chain.

\begin{figure}[t]
\centering
\includegraphics[width=0.39\textwidth,trim=0 0 0 70,clip]{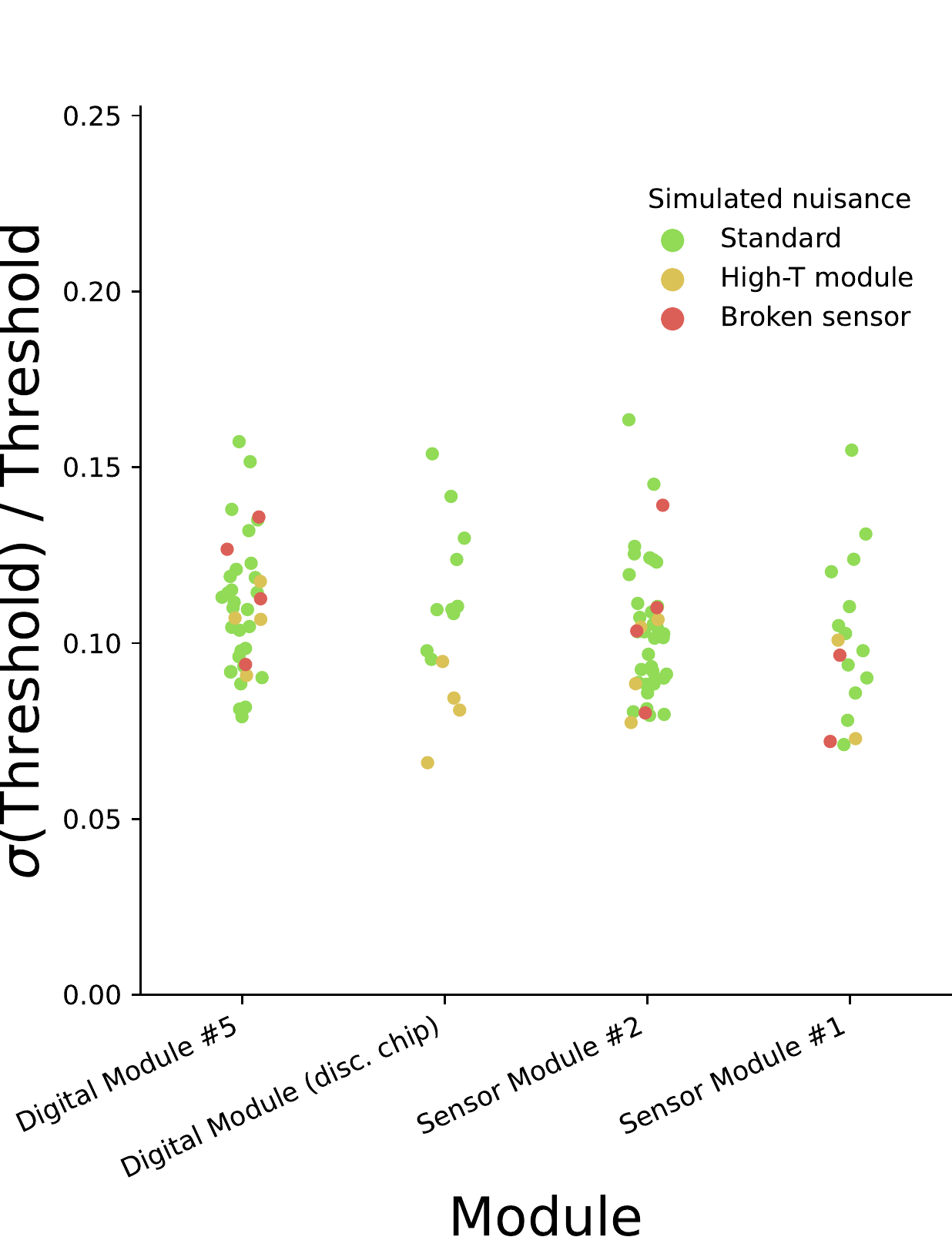} \hfill
\includegraphics[width=0.39\textwidth,trim=0 0 0 70,clip]{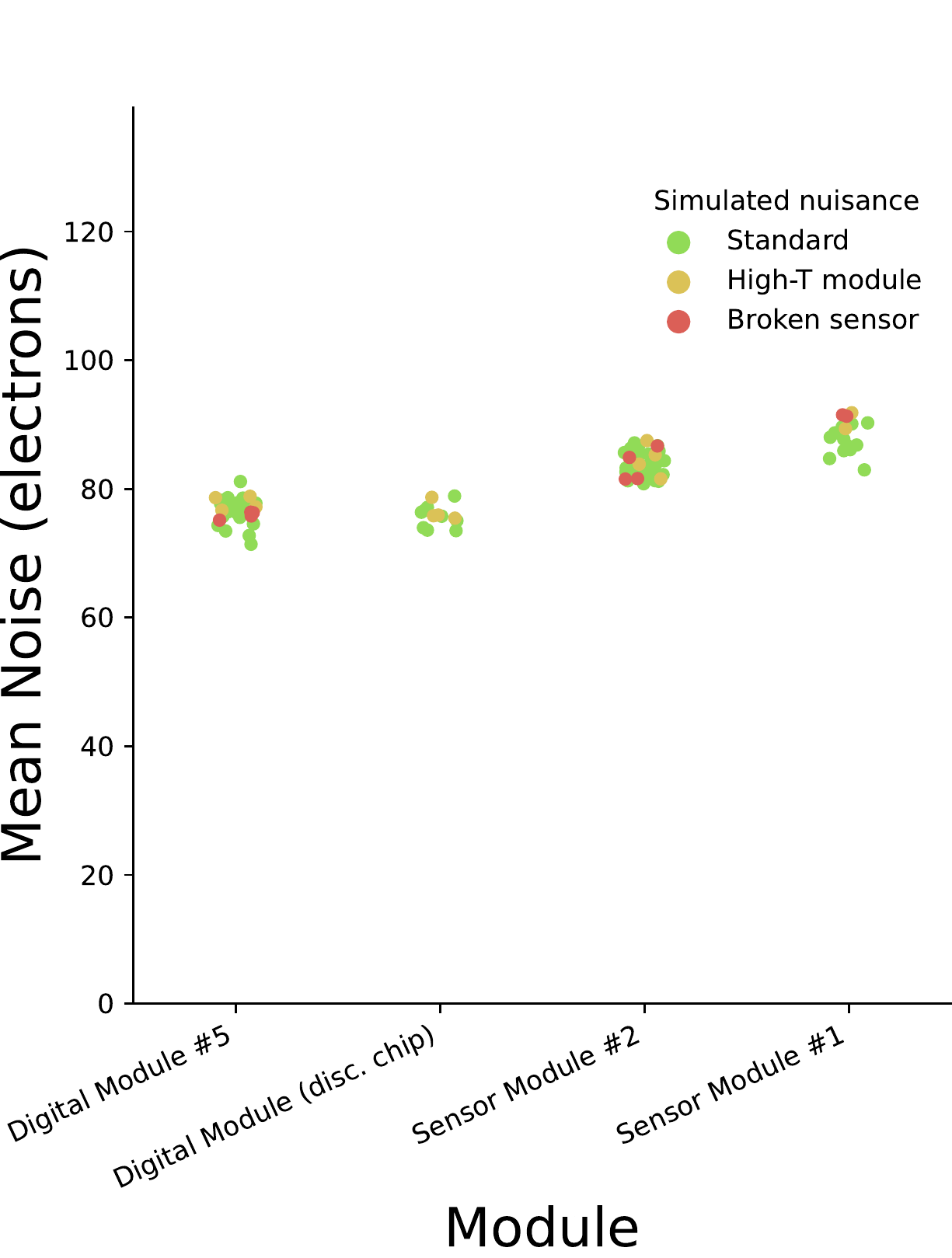}
\caption{Comparisons of the relative width of the threshold distribution (left) and the mean noise (right) between standard operation in standalone and serial powering operation (green) and two possible failure scenarios in the serial powering mode (yellow and red) as described in the text for different TEPX prototype modules ($x$-axis). Each point corresponds to a single measurement with a single ROC.}
\label{fig:Nuisances}
\end{figure}

The resilience of the SP chain against potential failures is tested in two scenarios. First, the middle one of the five fully operative modules in R1 is replaced by one that has only three out of four ROCs wire-bonded to the HDI. Consequently, the input current is shared between fewer chips, in turn causing increased chip temperatures. Second, a module with a broken silicon sensor, showing a high bias current, is inserted instead. In Figure~\ref{fig:Nuisances}, the relative width of the threshold distribution (left) and the mean noise (right) are compared between these two scenarios and standard SP operation. No deviation from the expected behavior is observed, indicating that the scenarios tested do not affect other modules in the serial powering chain.

\subsection{Signal Integrity Characterization}
\label{subsec:ber}

The signal integrity along the electrical TEPX readout chain is tested by performing electrical bit error rate (BER) tests using the PRBS7 pattern of the RD53A ROC. In particular, results for the longest data line in the TEPX system with a length of 492\,mm are reported. It corresponds to the fifth position in ring 1 (position R15) of the TEPX half-disk prototype, the leftmost position in R1 as shown in Figure~\ref{fig:DiskAndSP} (left).

Here, the BER is defined as the number of 32-bit frames with at least one bit error divided by the total number $N_{\rm{sent}}$ of frames sent. The nominal data transmission rate of $1.28\,\mathrm{Gb}/\mathrm{s}$ is used. In case no bit errors are recorded, a BER of $1/N_{\rm{sent}}$ is reported, corresponding to a conservative upper limit on the true BER. While measuring the BER on a given chip, all other ROCs in the SP chain transmit AURORA-encoded data.

\begin{figure}[t]
\centering
\includegraphics[width=0.44\textwidth]{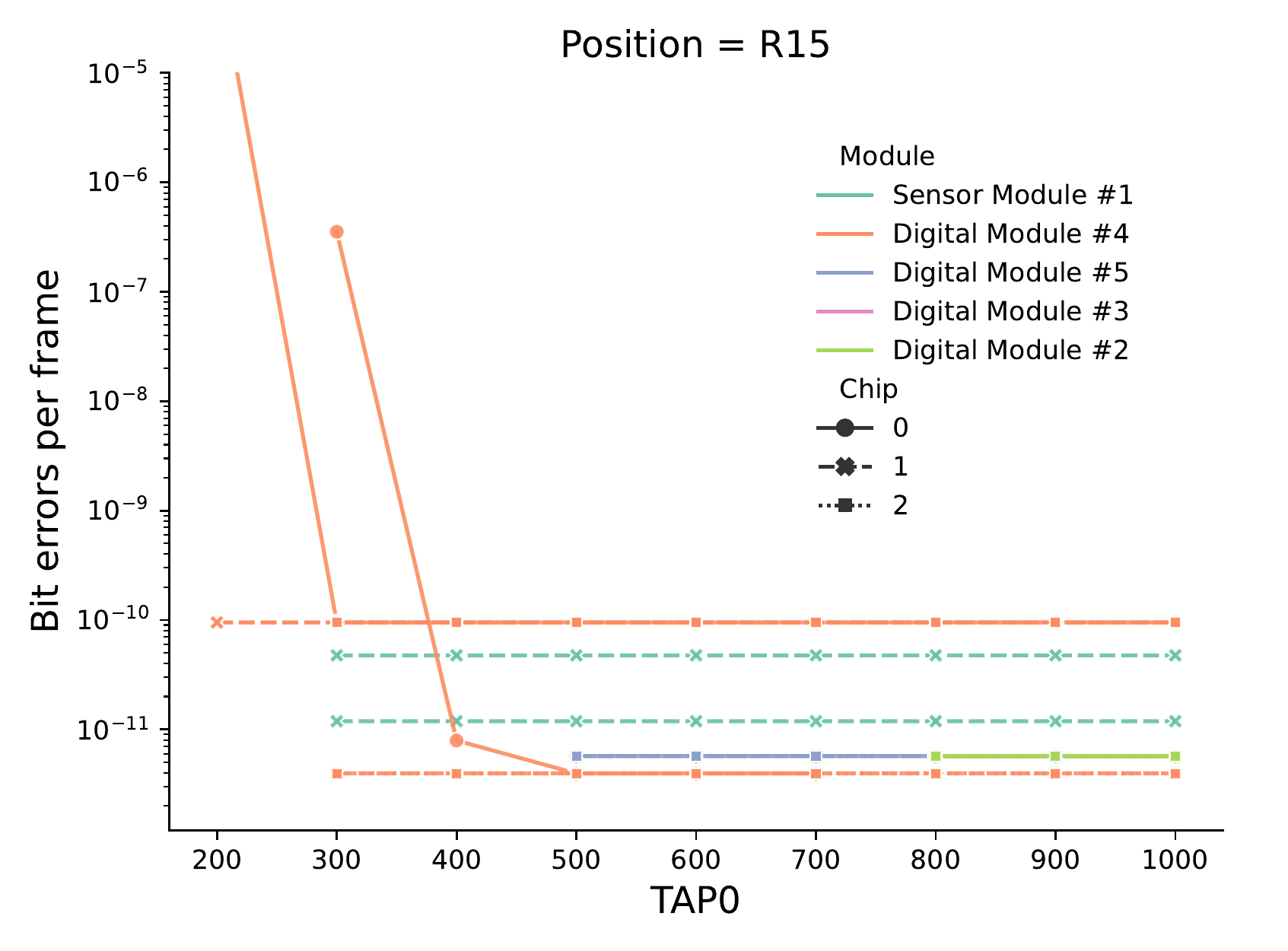} \hfill
\includegraphics[width=0.44\textwidth]{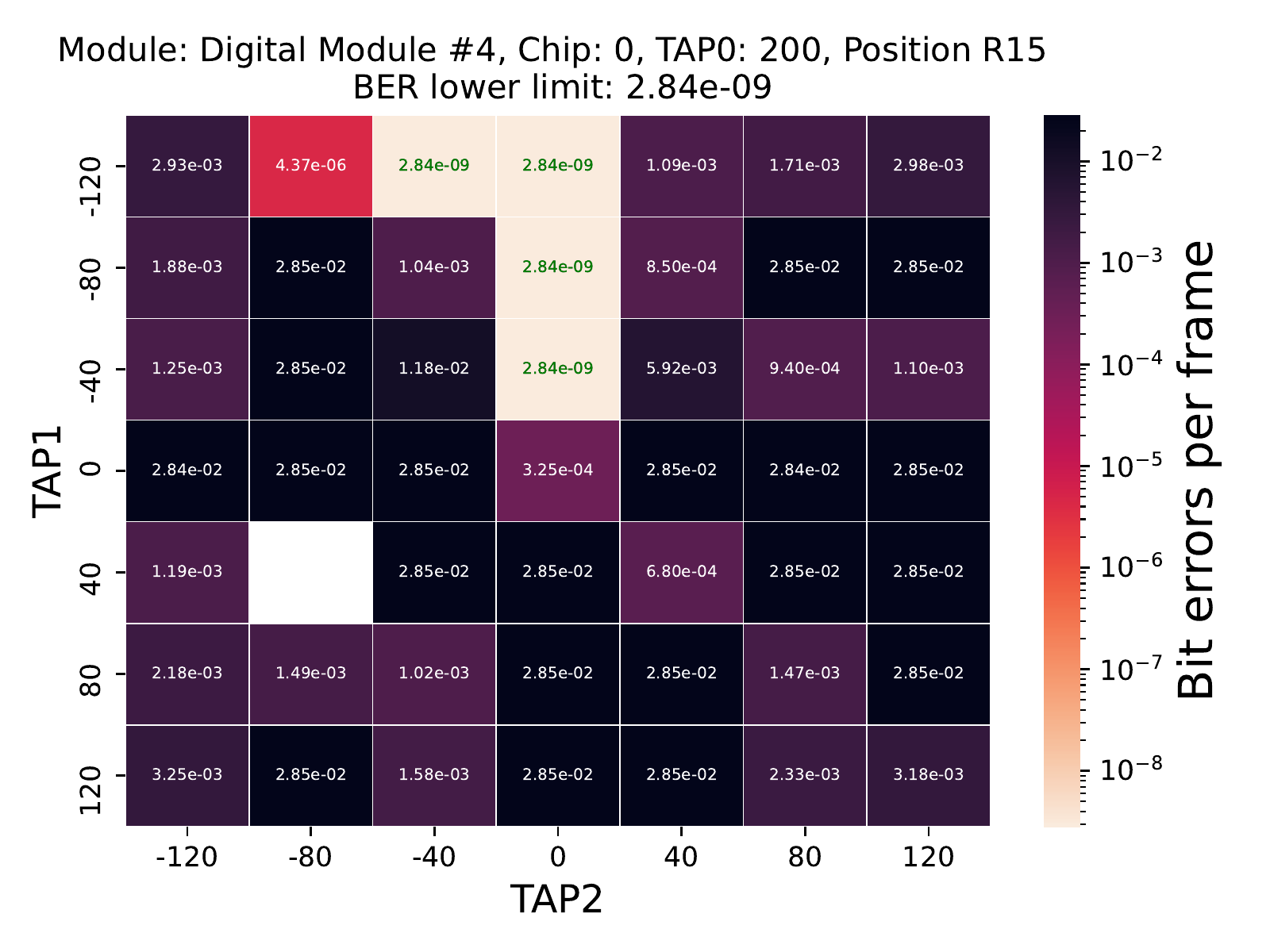}
\caption{Bit error rate as defined in the text as a function of TAP0 (left) and TAP1 and TAP2 (right).}
\label{fig:BER}
\end{figure}

First, the BER is measured for varying signal amplitudes, which are steered using the \textsc{CML\_TAP0\_BIAS} register (TAP0) of the RD53A ROC. TAP0 can take values between 0 and 1023, with the default value being 500. In Figure~\ref{fig:BER} (left), the BER as a function of TAP0 is shown for different ROCs on different TEPX modules in position R15. Very strong improvement of the BER is found with increasing signal amplitude for all modules and ROCs tested. It is possible to consistently achieve 0 bit errors at or already below the default RD53A setting, translating to a BER below $4\times 10^{-12}$ and leaving a considerable margin for future optimization.

In order to study the effect of first- and second-order pre-emphasis on the BER, very small signal amplitudes corresponding to $\rm{TAP0} = 200$ are used. Only under such conditions a non-zero BER could be measured. The first- and second-order pre-emphasis is steered by the \textsc{CML\_TAP1\_BIAS} and \textsc{CML\_TAP2\_BIAS} registers of the RD53A chip (TAP1 and TAP2), respectively. In Figure~\ref{fig:BER} (right), the BER is shown as a function of TAP1 and TAP2. An improvement of the signal integrity with inverted first-order pre-emphasis of the signal is observed, while no dependence on second-order pre-emphasis is found. Consequently, an optimized choice of the first-order pre-emphasis can be used to further improve the signal integrity in future operation.

Potential cross-talk between the data lines in the TEPX disk has been tested by measuring the BER for three different serializer output modes of the remaining ROCs in the SP chain.
Only a small improvement with respect to the AURORA mode is observed for the PRBS7 and GND modes, and only for the smallest signal amplitude corresponding to $\rm{TAP0}=200$ tested, which indicates insignificant cross-talk between data lanes in the SP chain.

\section{Conclusion}
\label{sec:conclusion}

The Tracker Endcap Pixel detector foreseen to be implemented during the upgrade of the CMS detector for the High-Luminosity phase of the LHC has been presented. The serial powering scheme and its influence on the pixel threshold tuning has been studied. No additional noise was found to be introduced by the serial powering chain and two potential failure scenarios were found to not affect other modules in the chain. The electrical signal integrity was studied by means of bit error rate tests, demonstrating excellent signal quality in the longest TEPX data line and leaving a promising margin for future optimization.

\bibliographystyle{JHEP}
\bibliography{bibliography}

\providecommand{\href}[2]{#2}\begingroup\raggedright\begin{thebibliography}{1}

\bibitem{Bruning:782076}
O.S.~Brüning~et al., \emph{{LHC Design Report}}, CERN (2004),
  \href{https://doi.org/10.5170/CERN-2004-003-V-1}{10.5170/CERN-2004-003-V-1}.

\bibitem{Evans:2008zzb}
L.~Evans and P.~Bryant, \emph{{LHC Machine}},
  \href{https://doi.org/10.1088/1748-0221/3/08/S08001}{\emph{JINST} {\bfseries
  3} (2008) S08001}.

\bibitem{Aberle:2749422}
O.~Aberle~et al., \emph{{High-Luminosity Large Hadron Collider (HL-LHC):
  Technical design report}}, CERN (2020),
  \href{https://doi.org/10.23731/CYRM-2020-0010}{10.23731/CYRM-2020-0010}.

\bibitem{CERN-LHCC-2017-009}
{CMS Collaboration}, \emph{{The Phase-2 Upgrade of the CMS Tracker}},  Tech.
  Rep. \href{https://cds.cern.ch/record/2272264}{CMS-TDR-014} (2017).

\bibitem{Garcia-Sciveres:2287593}
{RD53 Collaboration}, \emph{{The RD53A Integrated Circuit}},  Tech. Rep.
  \href{https://cds.cern.ch/record/2287593}{CERN-RD53-PUB-17-001} (2017).

\bibitem{Pesaresi:2015qfa}
M.~Pesaresi~et al., \emph{{The FC7 AMC for generic DAQ \& control applications
  in CMS}}, \href{https://doi.org/10.1088/1748-0221/10/03/C03036}{\emph{JINST}
  {\bfseries 10} (2015) C03036}.

\end{thebibliography}\endgroup

\end{document}